\documentclass[aps,prd,preprint,superscriptaddress]{revtex4-1}
\usepackage{amsmath,amssymb}
\usepackage[colorlinks=true,urlcolor=blue,citecolor=blue,linkcolor=blue]{hyperref}
\usepackage{graphicx}

\begin{document}

\title{Quantum simulation of dark energy candidates}

\author{Daniel Hartley}
\email[Corresponding author: ]{daniel.hartley@univie.ac.at}
\affiliation{Faculty of Physics, University of Vienna, Boltzmanngasse 5, 1090 Wien, Austria}
\author{Christian K{\"a}ding}
\affiliation{School of Physics \& Astronomy, University of Nottingham, University Park, Nottingham NG7 2RD, United Kingdom}
\author{Richard Howl}
\author{Ivette Fuentes}
\affiliation{School of Mathematical Sciences, University of Nottingham, University Park, Nottingham NG7 2RD, United Kingdom}

\date{\today}

\begin{abstract}
Additional scalar fields from scalar-tensor, modified gravity or higher dimensional theories beyond general relativity may account for dark energy and the accelerating expansion of the Universe. These theories have led to proposed models of screening mechanisms, such as chameleon and symmetron fields, to account for the tight experimental bounds on fifth-force searches. Cold atom systems have been very successfully used to constrain the parameters of these screening models, and may in future eliminate the interesting parameter space of some models entirely. In this paper, we show how to manipulate a Bose-Einstein condensate to simulate the effect of any scalar field model coupled conformally to the metric. We give explicit expressions for the simulation of various common models. This result may be useful for investigating the computationally challenging evolution of particles on a screened scalar field background, as well as for testing the metrology scheme of an upcoming detector proposal.
\end{abstract}

\maketitle

\section{Introduction}

General relativity (GR) is a tremendously successful theory and has made many accurate physical predictions. Nevertheless, often motivated by attempts to unify gravity with other forces of Nature (e.g. Kaluza-Klein theory \cite{Kaluza1921,Klein1926}), modifications to Einstein's theory of gravity have been proposed and investigated since its formulation. Amongst these modified theories of gravity, scalar-tensor theories (e.g. Brans-Dicke theory \cite{Brans1961}) are some of the most studied. This is partly due to the fact that coupling an additional scalar field to the metric tensor is one of the simplest ways to extend general relativity. Modified theories of gravity like $f(R)$-gravity (see \cite{Sotiriou2008} for a review) can additionally be shown to be equivalent to scalar-tensor theories, and higher dimensional theories (e.g. string theory) predict the existence of effective scalar field modes in 4-dimensional spacetime due to compactifications of the extra dimensions (see e.g. \cite{Wehus2002}).

Modifications of gravity gained even greater attention after the accelerated expansion of the Universe was discovered \cite{Perlmutter1998,Riess1998} and the puzzle of dark energy - the energy that supposedly drives this expansion - arose. Consequently, there have been several proposed explanations for the nature of dark energy based on scalar-tensor theories (see e.g. \cite{Clifton2011,Joyce2014} for an overview of models). Some of these models have already been ruled out by observations but others are still viable theories (see e.g. \cite{Ishak2018} for an overview of allowed and excluded models). The theories that are still candidates for explaining dark energy (and potentially other phenomena like dark matter - see e.g. \cite{Khoury2014, Copeland2016} for recent ideas) must be studied further and tested experimentally (e.g. as was done in 2017 with the observation of gravitational waves from a binary neutron star merger \cite{LIGO2017,Langlois2017}). In particular, understanding the properties and effects of the involved scalar fields is of utmost importance and finding hints of their existence would be a scientific breakthrough.

Many scalar-tensor theories assume the coupling between the metric tensor and these scalar fields to be via a conformal factor, as will be elaborated on in Section \ref{subsec:scalarintro}. Cold atom systems (e.g. atom interferometers \cite{Copeland2014,Jaffe2017}) are great tools for studying the effects of scalar fields with this particular type of coupling in experiments, and constraining these models. As we will show in this article, Bose-Einstein condensates (BECs) are an excellent means to simulate the behaviour of matter in spacetimes with a conformally coupling scalar field. In particular, such analogue simulations will be very useful for better understanding the behaviour of matter under the influence of scalar fields with non-linear behavior since their non-linearities can make computer simulations challenging.

A good analogy can be an invaluable tool in studying a complex or inaccessible system. From the first theoretical proposal to model a ``sonic horizon" with fluid flow in 1981 \cite{Unruh1981} to experiments ranging from water waves \cite{Weinfurtner2011,Euve2015,Euve2018} to BECs \cite{Lahav2010,Hung2013,Steinhauer2014}, nonlinear optical media \cite{Philbin2008,Vocke2018} and more, analogue gravity has become a widely varied approach for probing aspects of the physics of curved spacetime. BECs in particular have generated a rich catalogue of analogue models, as they are both relatively simple and manifestly quantum systems. This allows for the study of how non-classical properties, such as entanglement, are modified or generated by simulated gravitational fields. In a manifestly quantum system, we also have access to the powerful tools of quantum metrology.

Excerpts from this catalogue of analogue models include conformal Schwarzschild black holes \cite{Cropp2016,Dey2016}, rotating black holes \cite{Giacomelli2017}, Ba\~{n}ados-Teitelboim-Zanelli (BTZ) black holes \cite{Kaur2018}, Friedmann-Robertson-Walker geometries \cite{Barcelo2003,Fagnocchi2010}, inflation \cite{Fischer2004,Cha2017} and extensions to general relativity such as aether fields \cite{Cropp2016}. Three of us have extended this catalogue to include gravitational wave space-times \cite{Bravo2015,Hartley2018}. As an example, there has recently been an experimental implementation of a waterfall horizon leading to observation of density correlations across the horizon \cite{Steinhauer2016,deNova2018}, and controversy over their interpretation as entangled acoustic Hawking radiation \cite{Leonhardt2018}.

In our particular case, simulating the effect of conformally coupled screened scalar field models on matter in a covariant formalism will inform potential future work on the effect of screened scalar fields on particle creation, decoherence or other interesting quantum effects, as well as allowing us to test the metrological scheme of an upcoming proposal for a detector \cite{wip} constraining these models. The approach we take towards modelling our BEC and its phonons comes from quantum field theory in curved spacetime. It is manifestly covariant and thus allows us to introduce the background spacetime in a clear and natural way.

While much can be learned from the study of analogue systems, it should be stressed that analogue is of course not identity. Thus, no analogue can be used to falsify a theory or truly discover new effects and features. If some new or unexpected result presents itself in an analogue model, great care must be taken to distinguish real physical features of the system being modelled from artefacts of the analogue system. What analogue systems can teach us is how and where to look for features and effects of a theory or real system, as well as performing important self-consistency checks and demonstrating effects that are not reliant on precise and complete accuracy of the analogy. Testing the metrological scheme of a detector as mentioned above is an example of one such effect.

In Section \ref{subsec:scalarintro}, we introduce conformally coupled screened scalar field models. In Section \ref{subsec:acousticmetric}, we introduce the ``acoustic metric,'' a description of the evolution of phonons on a BEC as a scalar field on some curved background which depends both on the real background spacetime as well as the mean-field properties of the BEC.

In Section \ref{sec:sim}, we derive both the direct simulation of a conformally coupled screened scalar field model metric, as well as the simulation of the effect of such a metric on phonons on the BEC. For clarity, the first of these simulations (direct simulation) is a manipulation of the experimental parameters of the BEC to produce a situation in which the phonons on the BEC act as if they were a massless non-interacting boson field in a spacetime goverened by the conformal metric introduced in Section \ref{subsec:scalarintro}. The second of these simulations (effect simulation) involves deriving and then artificially reproducing the effect of a conformally coupled screened scalar field as if it were in the real background metric. As it turns out (and as will be shown in Section \ref{sec:sim}), these scenarios happen to be mathematically equivalent in this case. These two scenarios are not always equivalent, e.g. for gravitational waves in \cite{Hartley2018}. Note that neither of these scenarios is equivalent to simulating the evolution of the screened scalar field itself.

In Section \ref{sec:models}, we present some of the major conformally coupled screened scalar field models, comprising chameleons, symmetrons, dilatons, galileons and D-BIons, and give explicit examples for simulating these particular models. We conclude in Section \ref{sec:conclusion}.

\section{Background}
\label{sec:background}

\subsection{Conformally coupled screened scalar fields}
\label{subsec:scalarintro}

Theories like string theory or f(R)-gravity naturally predict the existence of additional scalar fields coupled to the metric tensor. Some of these fields can be part of possible explanations for the nature of dark energy in the form of models like quintessence (see e.g. \cite{Steinhardt2003}), or simply by having a non-vanishing vacuum expectation value which leads to a cosmological constant term. For example, a scalar field $\varphi$ with potential $V(\varphi) = \lambda\varphi^4 - \Lambda^3\varphi$ has a vacuum expectation value $\varphi_0 \neq 0$ and leads to a cosmological constant $\lambda\varphi_0^4 - \Lambda^3\varphi_0$ after splitting $\varphi = \varphi_0 + \delta\varphi$ into a constant background and a fluctation term.

The resulting modifications of general relativity (GR) are investigated in scalar-tensor theories of gravity \cite{Fujii2003}. Since an additional scalar field is coupled to the metric tensor in these theories, the Einstein-Hilbert action is modified correspondingly. For example, f(R)-gravity leads to a gravitational action
\begin{equation}
S_G = \frac{M_P^2}{2}\int d^4x\sqrt{-\tilde{g}}\Phi\tilde{R},\label{eq:fr_action}
\end{equation}
where $M_P$ is the reduced Planck mass and $\Phi$ is an additional scalar field. With $^\sim$ we denote a quantity in the Jordan frame which is one of infinitely many possible conformal frames \cite{Domenech2016} in which a theory of gravity can be formulated in. There are no explicit interactions between the scalar and the matter fields in this particular frame.

Besides the Jordan frame, it is common practice to make use of a second specific frame called the Einstein frame. In this particular frame the gravitational action takes on the canonical form of the Einstein-Hilbert action but is accompanied by some additional scalar field terms. For example, the f(R)-gravitational action (\ref{eq:fr_action}) is given in the Einstein frame by
\begin{equation}
S_G = \frac{M_P^2}{2}\int d^4x\sqrt{-g}\left(R+3\frac{\square\Phi}{\Phi}-\frac{9}{2}\left(\nabla\ln\Phi\right)^2\right),
\end{equation}
and a redefinition of $\Phi=:\exp\left[2\varphi/\sqrt{6}M_P\right]$ leads to a canonical kinetic term for the scalar field $\varphi$.

Besides serving as different mathematical formulations of the same theory, Jordan and Einstein frames also lead to different but ultimately equivalent interpretations of the same
physical phenomenon. In the Jordan frame formulation, Einstein's theory of gravity is modified in such a way that test particles follow different geodesics from those predicted in GR, while in the Einstein frame formulation, test particles still follow GR's geodesics but are also subject to a gravity-like fifth force of Nature carried by the additional scalar field $\varphi$. This hypothetical fifth force will be further discussed later in this subsection.

In order to go from one frame to another, a conformal transformation is applied such that Jordan frame metric $\tilde{g}_{\mu\nu}$ and Einstein frame metric $g_{\mu\nu}$ are related by
\begin{equation}
\tilde{g}_{\mu\nu}=A\left(\varphi\right)g_{\mu\nu}.\label{eq:frame_conversion}
\end{equation}
Using the frequently applied definition $A\left(\varphi\right)=:\exp\left[\zeta^2\left(\varphi\right)\right]$ and assuming $\left|\zeta\right|^2\ll1$, Eq. (\ref{eq:frame_conversion}) can be reduced to
\begin{equation}\label{eqn:conformaltrafo}
\tilde{g}_{\mu\nu}=\left(1+\zeta^2+\mathcal{O}\left(\zeta^4\right)\right)g_{\mu\nu}.
\end{equation}

All experiments and observations performed so far within our solar system have been in agreement with the predictions of GR (e.g. lunar laser ranging \cite{Dickey1994} and gravitational lensing measurements \cite{Bartelmann2010}). To be consistent with these experiments, any theory beyond GR must modify these predictions at most perturbatively, which is why we may assume $\left|\zeta\right|^2\ll1 $.

In this article we will refer to scalar fields coupling to the Einstein frame metric in this way as ``conformally coupling scalar fields''. This particular way of coupling a scalar field $\varphi$ to the gravitational metric can give rise to a coupling between $\varphi$ and ordinary matter fields via the trace of the energy momentum tensor $T^{\mu\nu}$. Such a coupling gives rise to a fifth force whose existence has not yet been confirmed by local observations \cite{Adelberger2009}. It should be stressed that this fifth force is generally not considered to be an explanation for the accelerated expansion of the Universe, but rather is a byproduct of some scalar field models used to explain dark energy. Assuming the existence of additional scalar fields, this apparent absence of a fifth force poses a conundrum as the scalar fields' coupling to matter requires it to be there. A simple solution would be to assume this coupling to be extremely weak, such that the force is universally unmeasurably weak. However, such a scenario would be a result of fine-tuning and of little to no observational interest.

A much more interesting solution is to assume that the scalar field is subject to a screening mechanism - a mechanism that suppresses the scalar field and its force in environments of high mass density but allows them to act with their full strength in vacuum. There are several models for such “screened scalar fields” with different types of screening mechanisms, some of which we will present in more detail in Section \ref{sec:models}. Screened scalar fields have been the subject of many different types of experiments in recent years and there are still ongoing efforts to constrain and detect them. An overview of experiments for some popular models can be found in \cite{Sakstein2016,Burrage2017}.

\subsection{Acoustic metric}

\label{subsec:acousticmetric}

A BEC is formed when many bosons are all collectively brought into the same state so that quantum features (such as long range order and wavefunction interference) determine the macroscopic behaviour of the boson collective. Practically, this is done by cooling a dilute gas of bosons such that the vast majority of the boson population is in the ground state. To simulate the effect of a screened scalar field on a BEC, we will describe the BEC as a barotropic, irrotational and inviscid fluid in a covariant formalism, following the approach of \cite{Bruschi2014,Hartley2018}. As in \cite{Hartley2018}, we emphasise that we are considering a regular BEC as those currently demonstrated in experiments so there are no ``relativistic'' features such as high speeds, large energies or strong coupling.

Following the above references, we describe the evolution of the total field operator $\hat{\Phi}$ of the BEC with the Lagrangian density
\begin{equation}
\mathcal{L}=-\sqrt{-g}\left\{ g^{\mu\nu}\partial_{\mu}\hat{\Phi}^{\dagger}\partial_{\nu}\hat{\Phi}+\left(\frac{m^{2}c^{2}}{\hbar^{2}}+V\right)\hat{\Phi}^{\dagger}\hat{\Phi}+U\left(\hat{\Phi}^{\dagger}\hat{\Phi},\lambda_{i}\right)\right\}, \label{eq:lagrangian}
\end{equation}
where $g_{\mu\nu}$ is the metric of the (in general curved) background spacetime, $V$ is the applied external potential and $U$ is the interaction potential defined as
\begin{equation}
U\left(\hat{\Phi}^{\dagger}\hat{\Phi},\lambda_{i}\right)=\frac{\lambda_{2}}{2!}\hat{\Phi}^{\dagger}\hat{\Phi}^{\dagger}\hat{\Phi}\hat{\Phi}+\frac{\lambda_{3}}{3!}\hat{\Phi}^{\dagger}\hat{\Phi}^{\dagger}\hat{\Phi}^{\dagger}\hat{\Phi}\hat{\Phi}\hat{\Phi}+\cdots.
\end{equation}
As usual, the $n$-th term of this expansion of $U$ corresponds to $n$-particle interactions. We will only consider the first order two-particle contact interaction term and neglect higher order terms, as is standard when describing simple BEC systems \cite{PitStrBEC}. For notational convenience, we drop the index on $\lambda$ and write
\begin{equation}\label{eq:interaction}
U\left(\hat{\Phi}^{\dagger}\hat{\Phi},\lambda_{i}\right)\approx\frac{\lambda}{2}\hat{\Phi}^{\dagger}\hat{\Phi}^{\dagger}\hat{\Phi}\hat{\Phi}.
\end{equation}
This interaction strength $\lambda$ can be related to first order to the s-wave scattering length $a$ by
\begin{equation}
\lambda=8\pi a.
\end{equation}
We then seperate the total field $\hat{\Phi}$ into the ``classical" mean field $\phi=\sqrt{\rho}e^{i\theta}$ and the quantum fluctuations $\hat{\psi}$ as
\begin{equation}
\hat{\Phi}=\phi\left(1+\hat{\psi}\right),
\end{equation}
where we have made the Bogoliubov approximation so $\left<\hat{\Phi}\right>\approx\phi$ and $\left<\hat{\psi}^\dagger\hat{\psi}\right>\ll\left|\phi\right|^2$. In the long wavelength limit, these fluctuations $\hat{\psi}$ take the form of massless non-interacting phonons. We can write the long wavelength limit explicitly as
\begin{equation}
\left|k\right|\ll\frac{\sqrt{2}}{\xi}\left(1+\frac{\hbar^{2}}{2m^{2}\xi^{2}u_{0}^{2}}\right)\min\left(1,\frac{mu_{0}\xi}{\sqrt{2}\hbar}\right)
\label{eq:longwave}
\end{equation}
where $k$ is the spatial frequency of a phononic excitation, $u$ is the flow velocity of the BEC defined as
\begin{equation}
u_{\mu}=\frac{\hbar}{m}\partial_{\mu}\theta,
\end{equation}
and $\xi$ is the ``healing length'' defined as
\begin{equation}
\xi=\frac{1}{\sqrt{\lambda\bar{\rho}}}
\end{equation}
where $\bar{\rho}$ is the average density.

The healing length can be physically motivated as follows: consider our field $\hat\Phi$ in a 1-D box trap of length $L$ with infinitely high walls. If the interaction strength vanishes (i.e. $\lambda=0$), then the ground state wave function has the familiar harmonic oscillator ground state shape \cite{PitStrBEC}
\begin{equation}
\left|\phi\right|^2 = \frac{2\bar{\rho}}{L}\sin^2\left[k\left(x+\frac{L}{2}\right)\right],
\end{equation}
where $k=\pi/L$. However, if $\lambda\ne0$ then the ground state wave function is approximately constant in space far from the boundaries. Close to the boundary $x=-L/2$, we have \cite{PitStrBEC}
\begin{equation}
\left|\phi\right|^2 \propto \tanh^2\left[\frac{x+L/2}{\sqrt{2}\xi}\right]
\end{equation}
so $\phi$ is approximately constant in space everywhere except an interval of the order $\xi$ near the boundaries. In the long wavelength limit of Eq. (\ref{eq:longwave}), we consider phonon wavelengths much longer than the healing length $\xi$ so for a box trap, the mean field density is approximately constant everywhere.

Taking the equations of motion of Eq. (\ref{eq:lagrangian}) and splitting the mean field $\phi$ and perturbations $\hat{\psi}$ as in \cite{Hartley2018} in $3+1$ dimensions, we arrive at a Klein-Gordon-like equation of the form
\begin{equation}
\frac{1}{\sqrt{-G}}\partial_{\mu}\sqrt{-G}G^{\mu\nu}\partial_{\nu}\hat{\psi}=0
\end{equation}
where $G_{\mu\nu}$ is a tensor acting like an effective ``acoustic metric'' with the form
\begin{equation}
G_{\mu\nu}=\frac{\rho c}{c_{s}}\left[g_{\mu\nu}+\left(1-\frac{c_{s}^{2}}{c^{2}}\right)\frac{v_{\mu}v_{\nu}}{c^{2}}\right].\label{eq:acousticmetric}
\end{equation}
In Eq. (\ref{eq:acousticmetric}), we have defined the normalised flow velocity as
\begin{equation}
v^{\mu}=\frac{c}{\left|u\right|}u^{\mu}
\end{equation}
and the scalar speed of sound as
\begin{equation}
c_{s}^{2}=\frac{c^{2}c_{0}^{2}}{c_{0}^{2}+\left|u\right|^{2}}\label{eq:c_s}
\end{equation}
where
\begin{equation}
c_{0}^{2}=\frac{\hbar^{2}}{2m^{2}}\rho\partial_{\rho}^{2}U\left(\rho,\lambda\right)=\frac{\hbar^{2}}{2m^{2}}\lambda\rho.
\end{equation}
Note that the definition of the flow velocity as the gradient of the phase imposes irrotationality, i.e.
\begin{equation}
\partial_{\mu}u_{\nu}=\partial_{\nu}u_{\mu}.
\end{equation}
The equations of motion of the mean field $\phi$ can be split into real and imaginary components resulting in a continuity equation
\begin{equation}
\nabla_{\mu}\left(\rho u^{\mu}\right)=0\label{eq:continuity}
\end{equation}
and an equation directly relating bulk field properties with potentials:
\begin{equation}
\left|u\right|^{2}=c^{2}+\frac{\hbar^{2}}{m^{2}}\left\{ V+\partial_{\rho}U\left(\rho,\lambda\right)-\frac{\nabla_{\mu}\nabla^{\mu}\sqrt{\rho}}{\sqrt{\rho}}\right\} .\label{eq:unorm}
\end{equation}
The continuity equation can also be derived directly from the Lagrangian as the conserved current associated to the global $U\left(1\right)$ phase symmetry of the total field.

It should be noted that this approach is more general and includes the effects of the background space-time in a more natural way than standard approaches in BEC analogue gravity. Starting from non-relativistic fluid equations or the Gross-Pitaevskii equation, an acoustic metric with the form
\begin{equation}\label{eq:nonrelacousticmetric}
G_{\mu\nu}\propto\begin{pmatrix}-\left(c_{s}^{2}-v^{2}\right)/c^{2} & -v_{i}/c\\-v_{j}/c & \delta_{ij}\end{pmatrix}
\end{equation}
can be derived, where $c_s$ is the speed of sound as above but $v$ is the 3-flow velocity (see \cite{Barcelo2005} and references therein). This model has been used for analogue gravity with both water waves and BECs, but is not a complete picture and notably omits any effects of the background space-time.

Eq. (\ref{eq:nonrelacousticmetric}) can be recovered from Eq. (\ref{eq:acousticmetric}) in the completely non-relativistic flat spacetime limit. A more complete model similar to that in Eq. (\ref{eq:acousticmetric}) has been derived in a relativistic field theory context in \cite{Fagnocchi2010} and from relativistic fluid equations in \cite{Visser2010}, but these both consider the background metric to be the flat Minkowski metric. Eq. (\ref{eq:acousticmetric}) has been derived in \cite{Bruschi2014,Hartley2018} and includes the effect of a (in general) curved background space-time metric in a natural way through a covariant formalism. This allows us to include any effects of the background metric arising at low (non-relativistic) energies when taking the non-relativistic limit to model real experiments.

\subsubsection{Non-relativistic limits}

As previously mentioned, we are not considering a BEC with relativistic energies, speeds or coupling strength. Hence, it is worth considering approximations that can be made with the definitions and results of Section \ref{subsec:acousticmetric} in the low energy limit.

Replacing the total field $\hat{\Phi}$ with a lower energy field
\begin{equation}
\hat{\Phi}=\hat{\Phi}_{NR}e^{imc^{2}t/\hbar}
\end{equation}
and considering the background spacetime to be flat results in the standard time dependent Gross-Pitaevskii equation (see e.g. \cite{PitStrBEC})
\begin{equation}
i\hbar\partial_t\hat{\Phi}_{NR}=\left[-\frac{\hbar^2}{2m}\nabla^2+V_{NR}+g_{NR}\hat{\Phi}^\dagger_{NR}\hat{\Phi}_{NR}\right]\hat{\Phi}_{NR}
\end{equation}
where the external potential and coupling strength are defined as
\begin{equation}
V_{NR}=\frac{\hbar^{2}}{2m}V\,,\,g_{NR}=\frac{\hbar^{2}}{2m}\lambda.
\end{equation}
When the external potential $V$ and interaction strength $\lambda$ are constant in time, the temporal frequency of the solutions to the Gross-Pitaevskii equation is the ground state energy, generally referred to as the normalised chemical potential $\mu/\hbar$. Thus, we can write the normalisation of the flow velocity $\left|u\right|$ as
\begin{equation}\label{eq:unorm2}
\left|u\right|=c+\frac{\mu}{mc}.
\end{equation}
Estimating a typical chemical potential as $\mu/\hbar\sim10^{3}Hz$ in a rubidium BEC (see e.g. \cite{Shammass2012} for typical experimental parameters in BEC analogue gravity), we have
\begin{equation}
\left|u\right|^{2}-c^{2}\sim10^{-6} m^2s^{-2}.
\end{equation}
In a harmonic trap where
\begin{equation}
V_{NR}\sim\frac{1}{2}m\omega_{i}^{2}x_{i}^{2}
\end{equation}
with a trapping frequency of $\omega\sim10^{2}Hz$, a $100\mu m$ wide cloud has a maximum potential of
\begin{equation}
\frac{\hbar^{2}}{m^{2}}V\sim\omega^{2}x^{2}\sim10^{-4} m^2s^{-2}.
\end{equation}
Hence, the contribution of the chemical potential in Eq. (\ref{eq:unorm}) through Eq. (\ref{eq:unorm2}) is two orders of magnitude below that of the applied potentials and thus can be neglected in this equation.

We can make similar approximations in the definition of $c_{s}$ in Eq. (\ref{eq:c_s}). Estimating the scattering length of rubidium as $a\sim100a_{0}$ \cite{Shammass2012} where $a_{0}$ is the Bohr radius in our approximately Gaussian cloud of width $100\mu m$ with $10^{6}$ atoms, $c_0\sim10^{-3}m/s$ so we can say $c_{0}^{2}\ll c^{2}$ with confidence. Thus, with the approximations $\mu\ll mc^{2}$ and $c_{0}^{2}\ll c^{2}$ we can say
\begin{equation}
c_{s}^{2}\approx c_{0}^{2}-\frac{c_{0}^{2}}{c^{2}}\left(c_{0}^{2}+\frac{\mu}{mc^{2}}\right)\approx c_{0}^{2}.
\end{equation}

\subsubsection{Defining flat space}

We define a ``flat'' acoustic metric for the phonons in the absence of any simulated fields as the acoustic metric in coordinates with the following conditions:
\begin{enumerate}
\item The background spacetime is flat, i.e. $g_{\mu\nu}=\eta_{\mu\nu}=\text{diag}\left(-1,1,1,1\right)$,
\item No flows, i.e. $v_{i}=0$,
\item Static density, i.e. $\rho_{0}=\rho_{0}\left(\boldsymbol{x}\right)\Leftrightarrow\partial_t\rho_{0}=0$, and
\item Unperturbed interaction strength $\lambda$, so $\partial_t\lambda=\partial_x\lambda=0$.
\end{enumerate}
Note that $\partial_t\rho=\partial_t\lambda=0$ and the approximation $c_s^2\approx c_0^2$ imply that $\partial_t c_s=0$. We assume in every case that the ``unperturbed'' bulk properties can be described in this way. The density $\rho_{0}$ is in general a function of space, but can also be constant by careful choice of trapping potentials as explained above in Section \ref{subsec:acousticmetric}.

The acoustic metric in $n+1$ dimensions then has the form
\begin{equation}
G_{\mu\nu}^{0}=\frac{\rho_{0}c}{c_{s0}}\begin{pmatrix}-c_{s0}^{2}/c^{2}\\
 & \mathbb{I}_{n}
\end{pmatrix}
\end{equation}
where $\mathbb{I}_{n}$ is the $n\times n$ identity matrix. If the speed of sound is also constant in space, then a simple time rescaling results in a metric trivially conformal to the $n+1$-dimensional Minkowski metric. Note that the ``no flows'' restriction implies that the phase of the mean field is a function of
time only.

\section{General simulation}
\label{sec:sim}

In this section, we present two results; the first in Section \ref{subsec:directsim} is a direct simulation of the Jordan frame metric in Eq. (\ref{eqn:conformaltrafo}) and the second in Section \ref{subsec:effectsim} is a simulation of the effect of a conformally coupled screened scalar field.

As explained in the Introduction, the direct simulation has the phonon field behaving as a massless non-interacting boson field on a background with a conformally coupled screened scalar field, whereas the effect simulation recreates the equations of motion for the phonons with a real conformally coupled screened scalar field in the background metric affecting the atoms of the BEC, with scalar field parameters chosen by an experimentalist for the simulation. It bears repeating that neither of these simulations is a simulation of the evolution of the conformally coupled screened scalar field.

\subsection{Direct simulation}

\label{subsec:directsim}

We first wish to directly simulate a Jordan frame metric, so
\begin{equation}
G_{\mu\nu}^{\left(sim\right)}=\left(1+\zeta^{2}\right)G_{\mu\nu}^{0}.
\end{equation}
Explicitly, this is
\begin{equation}
\frac{\rho c}{c_{s}}\begin{pmatrix}-c_{s}^{2}/c^{2}\\
 & \mathbb{I}_{n}
\end{pmatrix}=\left(1+\zeta^{2}\right)\frac{\rho_{0}c}{c_{s0}}\begin{pmatrix}-c_{s0}^{2}/c^{2}\\
 & \mathbb{I}_{n}
\end{pmatrix}.\label{eq:directsim}
\end{equation}
To preserve the structure of the metric, we must have $c_{s}=c_{s0}$. It is then clear that we must also have
\begin{equation}
\rho=\rho_{0}\left(1+\zeta^{2}\right).\label{eq:directrho}
\end{equation}
It is not possible to simultaneously satisfy all of the equations and conditions given above and in Section \ref{subsec:acousticmetric} in $3+1$ dimensions if $\zeta^{2}\left(\varphi\right)$ is a completely general function of time and space: if we write the bulk phase as
\begin{equation}
\theta\left(t\right)=\frac{mc^2}{\hbar}t+\vartheta\left(t\right)
\end{equation}
and require $\hbar\partial_t\vartheta\ll mc^2$ (non-relativistic limit) then Eq. (\ref{eq:continuity}) becomes
\begin{equation}
\partial_{t}\left(\rho\left(t,\boldsymbol{x}\right)\left[1+\frac{\hbar}{mc^{2}}\partial_{t}\vartheta\left(t\right)\right]\right)=0.\label{eq:nosolution}
\end{equation}
This does not have a general solution for $\vartheta\left(t\right)$ given some arbitrary $\rho\left(t,\boldsymbol{x}\right)$. As an example, consider
\begin{equation}
\rho\left(t,\boldsymbol{x}\right)=\rho_0\left(y,z\right)\exp\left[-\frac{x^2}{\left(\sigma+\varsigma t\right)^2}\right],
\end{equation}
i.e. a Gaussian expanding linearly in $x$ at a constant rate $\varsigma$. Substituting this into Eq. (\ref{eq:nosolution}) and rearranging terms, we find
\begin{equation}
2\varsigma x^{2}=-\left(\sigma+\varsigma t\right)^{3}\left[\frac{\hbar\partial_{t}^{2}\vartheta}{mc^{2}+\hbar\partial_{t}\vartheta}\right].
\end{equation}
This clearly has no solution for non-zero $\varsigma$ if $\vartheta$ is to be a function of time only. Note also that Eq. (\ref{eq:nosolution}) in the non-relativistic limit (which is representative of all real experiments) requires the density to be a function of space only. Eq. (\ref{eq:nosolution}) has a general solution if the density is only a function of space; namely $\vartheta = \mu t$ where $\mu$ is the chemical potential.

In the case where $\partial_{t}\zeta^{2}=0$ i.e. $\zeta^{2}$ is only a function of space or is constant, we can fulfill the continuity equation (Eq. \ref{eq:continuity}) to first order without modifying the chemical potential. From the definition of $c_{s}$, we can see that all of the above requirements can only be simultaneously fulfilled if the interaction strength is not fixed. It is well known that the interaction strength in a BEC can be modulated with an external magnetic field around a Feshbach resonance (see for example \cite{Schneider2012}).

To implement the density profile $\rho=\rho_0\left(1+\zeta^2\right)$ with an unchanged speed of sound, with Eqs. (\ref{eq:continuity}) and (\ref{eq:unorm}) we conclude that the external and interaction potentials should be modulated as
\begin{equation}
\lambda=\lambda_{0}\left(1-\zeta^{2}\right),
\end{equation}
\begin{equation}
\begin{split}V & =V_{0}+\left[\frac{\square\sqrt{\rho}}{\sqrt{\rho}}-\frac{\square\sqrt{\rho_{0}}}{\sqrt{\rho_{0}}}\right]\\
 & =V_{0}+\left[\frac{\partial^{i}\sqrt{\rho_{0}}}{\sqrt{\rho_{0}}}\left(\partial_{i}\zeta^{2}\right)+\frac{1}{2}\nabla^2 \zeta^{2}\right]
\end{split}
\end{equation}
in Cartesian coordinates, where $i$ runs over spatial indices and $\nabla^2$ is the Laplacian, which can be easily calculated for any specific form of the initial density $\rho_{0}$. The perturbation to $V$ captures the change in the shape of $\rho$ induced by the perturbation $\zeta^{2}$; if $\zeta^{2}$ doesn't vary in space, then the change in the interaction strength expands or contracts the BEC without modifying $c_{s}$ so no extra change in $V$ is necessary.

\subsection{Effect simulation}

\label{subsec:effectsim}

We now want to simulate the acoustic metric as seen by the phonons in the influence of some scalar field model. The conformal factor in the Jordan metric in Eq. (\ref{eqn:conformaltrafo}) enters the acoustic metric both directly and indirectly. The normalised flow velocity is normalised to $c^{2}$, so
\begin{equation}
-\tilde{g}^{\mu\nu}v_{\mu}v_{\nu}=-\tilde{g}^{00}v_{0}^{2}=c^{2}\implies v_{0}=c\left(1+\frac{1}{2}\zeta^{2}\right)
\end{equation}
Hence, we want
\begin{equation}
G_{\mu\nu}^{\left(sim\right)}\rightarrow G_{\mu\nu}^{\left(\phi\right)}=\frac{\rho_{0}c}{c_{s0}}\left(\tilde{g}_{\mu\nu}+\left(1-\frac{c_{s0}^{2}}{c^{2}}\right)\left(1+\zeta^{2}\right)\delta_{\mu\nu}^{00}\right).
\end{equation}
Expanding this expression, we have
\begin{equation}
\left(1+\zeta^{2}\right)\frac{\rho_{0}c}{c_{s0}}\begin{pmatrix}-c_{s0}^{2}/c^{2}\\
 & \mathbb{I}_{3}
\end{pmatrix}=\frac{\rho c}{c_{s}}\begin{pmatrix}-c_{s}^{2}/c^{2}\\
 & \mathbb{I}_{3}
\end{pmatrix}.\label{eq:effectsim}
\end{equation}
As this is identical to Eq. (\ref{eq:directsim}), the results of Section \ref{subsec:directsim} apply to this case as well. Note that this is not a given result for any metric to be simulated; e.g. the two types of simulation differ for a gravitational wave metric \cite{Hartley2018}.

\subsection{1+1 dimensions}

The reduction of these results to an effective $1+1$ dimensional condensate is not experimentally interesting as the phonons, in the low wavelength limit, behave as massless non-interacting Klein-Gordon bosons and thus their evolution is conformally invariant. Effects of any conformally coupled screened scalar field can only be seen here in the full $3+1$ dimensional analysis.

However, introducing an additional disformal coupling, such that the Jordan frame metric in Eq. (\ref{eqn:conformaltrafo}) takes on the form \cite{Bekenstein1992}
\begin{equation}
\tilde{g}_{\mu\nu}=\left(1+\zeta^2\left(\varphi\right)\right)g_{\mu\nu} + B\left(\varphi,\partial\varphi\right)\partial_\mu\varphi\partial_\nu\varphi,
\end{equation}
allows the scalar field to also couple to massless particles. $B$ is the disformal coupling parameter and can be choosen to be constant or a function of $\varphi$, $\partial\varphi$ depending on the model. Consequently, disformally coupling scalar fields could be treated when only considering an effective $1+1$ dimensional condensate, but this goes beyond the scope of this article.

\section{Models and examples}
\label{sec:models}

We will now give examples of conformally coupling scalar fields that are commonly used in modified theories of gravity and whose effects could be simulated as described in Section \ref{sec:sim} (see \cite{Clifton2011,Joyce2014,Koyama2015,Bull2015} for overviews of modern modified gravity theories and screened scalar fields). To give a simple but commonly used example, we consider a situation in which the screening for each field is sourced by a static homogeneous sphere of radius $R$ and give the resulting field profile outside the source. Every type of scalar field presented below has some individual type of screening mechanism. A screening mechanism allows a fifth force carried by a field to generally be strong (e.g. relative to gravity) but heavily suppressed in special circumstances (e.g. in the relatively dense environment within our solar system and, crucially, experimentally accessible space on and near the Earth).

For each presented scalar field (denoted by $\varphi$) we will give the Lagrangian density $\mathcal{L}_\varphi$ which specifies the model. With this we can define an action $S_\varphi := \int d^4x \mathcal{L}_\varphi$, such that the effective action of our four-dimensional Universe is schematically given in the Einstein frame by
\begin{eqnarray}
S_\text{4DUniverse} = S_\text{SM} + S_\text{Gravity} + S_\varphi 
\end{eqnarray}
with $S_\text{SM}$ denoting the action of the Standard Model of particle physics \cite{Woithe2017} and $S_\text{Gravity}$ the gravitational action. A more concrete description of how the gravitational action and the scalar field action are connected can be found in the form of Horndeski theory \cite{Horndeski1974}, which is the most general scalar tensor theory that gives rise to second order equations of motion.

The metric used in defining these models is the flat Minkowski metric. While any deviation of the real background metric from the Minkowski metric will have an effect on the form of these conformally coupled screened scalar field models, these deviations are observationally irrelevant. As pointed out in Section \ref{subsec:scalarintro}, the effect of these scalar field models must be a perturbation on GR to fit current experiments. Any deviation of the background metric from the Minkowski metric is therefore a perturbation of a perturbation and thus unmeasurable practically and unimportant to first order.

\subsection{Chameleon}

The chameleon scalar field model was first introduced in \cite{Khoury20032,Khoury2003} and deals with a screened scalar field $\varphi$ whose non-vanishing effective mass $m_\varphi$ is dependent on the environmental density. As its animal counterpart is adaptive to the colour of its surrounding, the chameleon field adapts its mass to the environment - a denser environment leads to a heavier chameleon mass.

\begin{figure}[tb]
\centering
\includegraphics[scale=0.5]{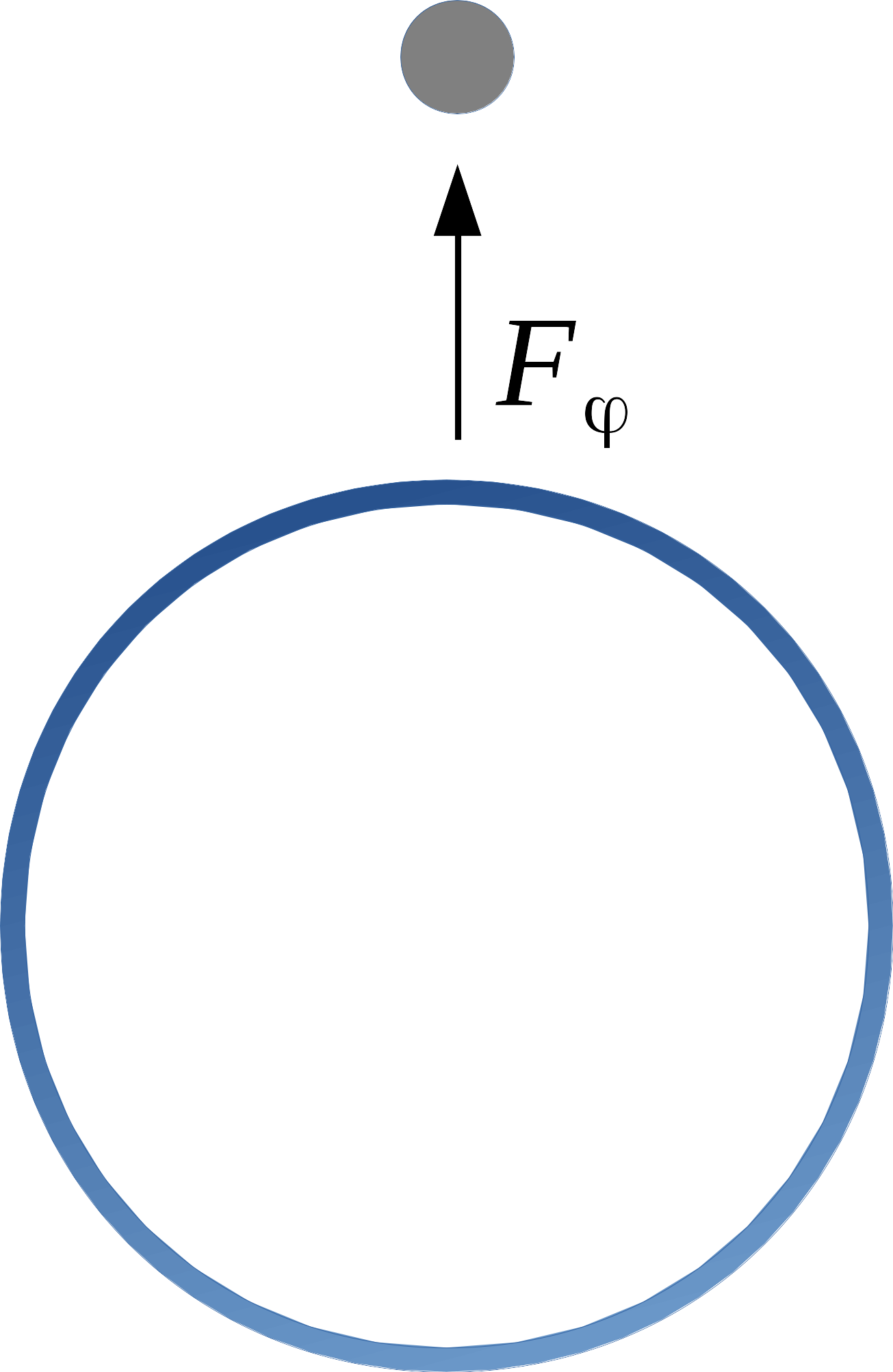}
\caption{Thin-shell effect: Only a thin shell of mass contributes to the chameleon force $F_\varphi$ sourced by a homgeneous massive object.}
\label{Fig:ChameleonForce}
\end{figure}

Since the chameleon force $F_\varphi(r)$ depends on the radius $r$ away from a chameleon source in the same way as a Yukawa potential \cite{Yukawa1935} (see also e.g. \cite{Peskin1995})
\begin{eqnarray}
F_\varphi(r) \sim \exp\left[-m_\varphi r c/ \hbar\right]/r, 
\end{eqnarray}
a heavy chameleon carries a shorter ranged force than a light one (the force is ``Yukawa suppressed''). This gives rise to the thin-shell effect which allows only the thin outermost layer of mass of an object to effectively contribute to the chameleon force (see Figure \ref{Fig:ChameleonForce}). Consequently, the chameleon fifth force coming from objects in our solar system is screened.

Nevertheless, there have been successful attempts to constrain the chameleon parameter space with Earth-based experiments since it is possible to create situations in which the field reaches its unscreened regime in the vacuum of a vacuum chamber (see \cite{Sakstein2016,Burrage2017} for an overview). This is also made possible by the thin-shell effect. More precisely, if the wall of a vacuum chamber is thick enough, the outside world will not contribute to the fifth force which allows it to be relatively unscreened.

The chameleon model is described by the Lagrangian density \cite{Khoury2003}
\begin{equation}\label{eqn:ChameleonLagrangian}
\mathcal{L}_\varphi=-\frac{1}{2}\left(\partial\varphi\right)^2-\frac{\Lambda^{4+n}}{\varphi^n}-\frac{\varphi}{M}\rho.
\end{equation}
where $n \in \mathbb{Z}^+ \cup \left\{x: -1 < x < 0 \right\} \cup 2\mathbb{Z}^-\backslash \left\{-2\right\}$ distinguishes between different chameleon models and $\rho$ is the density of matter. This matter coupling term arises from the coupling between $\varphi$ and $T^\mu_{~\mu}$ discussed in Section \ref{subsec:scalarintro}, under the assumption that matter can be approximated as a perfect fluid with negligible pressure. $\Lambda$ and $M$ determine the strength of the self-interaction and the chameleon-matter coupling, respectively.

\begin{figure}[tb]
\centering
\includegraphics[scale=0.1]{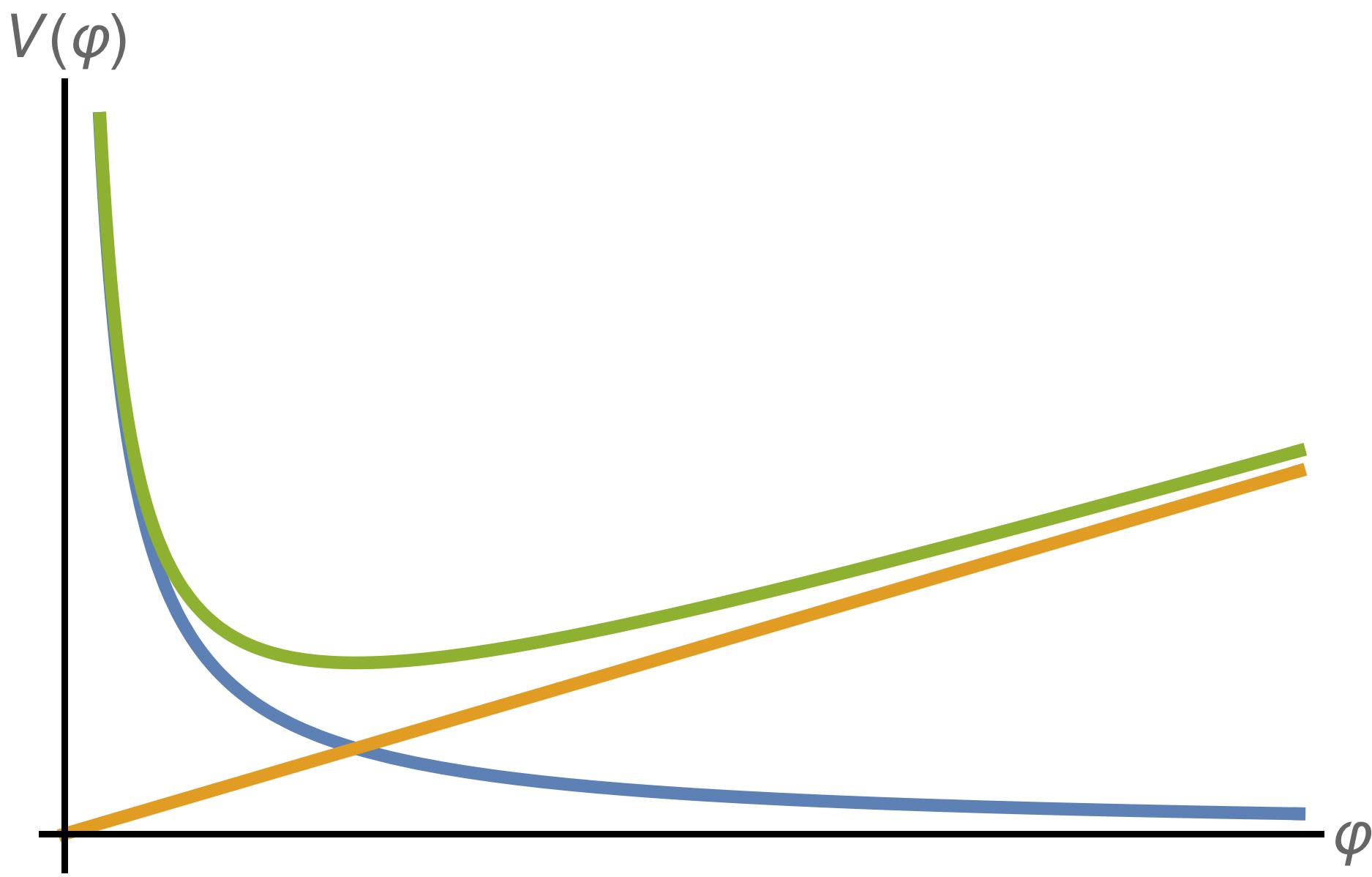}
\includegraphics[scale=0.1]{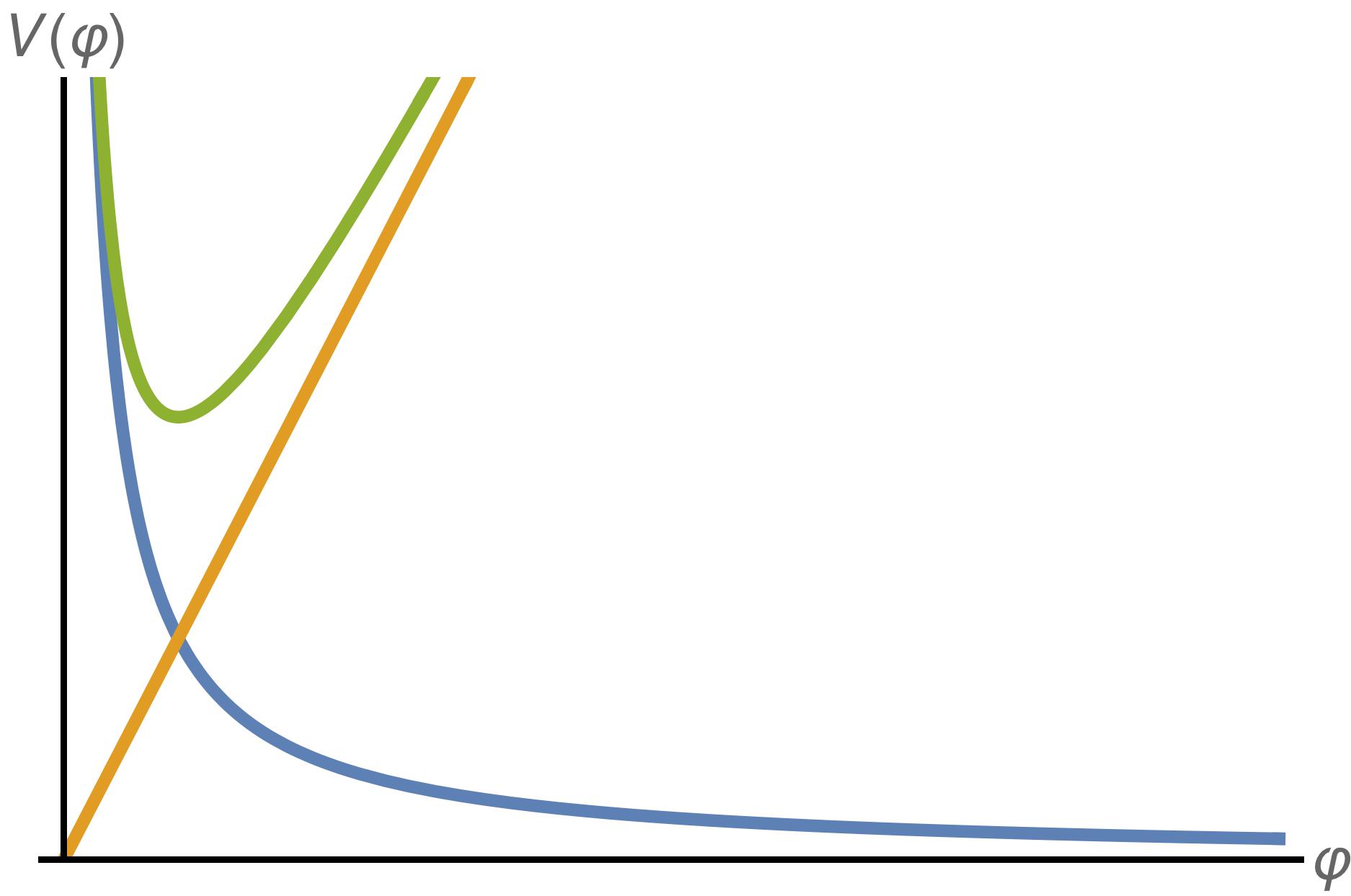}
\caption{The blue line represents the self-interaction potential of the chameleon (for $n>0$), the orange line depicts the interaction potential of the chameleon with matter and the green line 
is the effective potential given by the sum of those two. In contrast to the two potentials alone, the effective potential has a minimum which allows the chameleon to have a non-vanishing mass. 
The left (right) figure represents the case of a low (high) environmental mass density.}
\label{Fig:ChameleonPotential}
\end{figure}

The sum of the two final terms in Eq. (\ref{eqn:ChameleonLagrangian}) results in an effective potential with a local mininimum (see Figure \ref{Fig:ChameleonPotential}) and therefore in a non-vanishing, $\rho$-dependent chameleon mass. The exact value of this resulting mass is also controlled by $n$, $\Lambda$ and $M$.

Here $\zeta^2$ appearing in Eq. (\ref{eqn:conformaltrafo}) is given by $\zeta^2\left(\varphi\right)=\varphi/M$ and the field profile outside a static spherically symmetric source is given by \cite{Khoury2003}
\begin{equation}
\varphi\left(r\right) = -\left(\varphi_\infty-\varphi_{obj.}\right)\frac{R}{r}e^{-m_{\varphi,\infty}rc/\hbar}+\varphi_\infty,
\end{equation}
with $\varphi_{obj.}$ being the value of the chameleon inside the source. If the source is screened, $\varphi_{obj.}$ is actually the minimum of the chameleon within the source. A quantity which is labelled with $\infty$ is given in terms of $\rho_\infty$ which describes the mass density of the environment surrounding the object sourcing the chameleon (e.g. it would represent the density of the vacuum around a massive sphere in a vacuum chamber). The use of the label $\infty$ stems from the idealistic case of having an infinitely extended environment in which $\varphi(r) \to \varphi_\infty$ for $r \to \infty$.

To simulate a chameleon field Jordan frame metric, the external and interaction potentials as defined in Eqs. (\ref{eq:lagrangian}) and (\ref{eq:interaction}) must be
\begin{equation}
V\left(r\right)=V_0+\left[\frac{\partial_{r}\sqrt{\rho_{0}}}{\sqrt{\rho_{0}}}\left(\frac{1}{r}+\frac{m_{\varphi,\infty}c}{\hbar}\right)-\frac{m_{\varphi,\infty}^{2}c^{2}}{2\hbar^{2}}\right]\frac{\varphi_{\infty}-\varphi\left(r\right)}{M},
\end{equation}
\begin{equation}
\lambda\left(r\right)=\lambda_{0}\left(1-\frac{\varphi\left(r\right)}{M}\right).
\end{equation}

\subsection{Symmetron}

The symmetron is another commonly studied screened scalar field model and was first described in \cite{Dehnen1992, Gessner1992, Damour1994, Pietroni2005, Olive2008, Brax2010} and introduced with its current name in \cite{Hinterbichler2010,Hinterbichler2011}. As in the chameleon case, its screening mechanism is based on a coupling to matter which is dependent on the environmental density $\rho$. However, instead of being screened by having a heavy mass, the symmetron decouples from matter in dense environments.

\begin{figure}[tb]
\begin{center}
\includegraphics[scale=0.1]{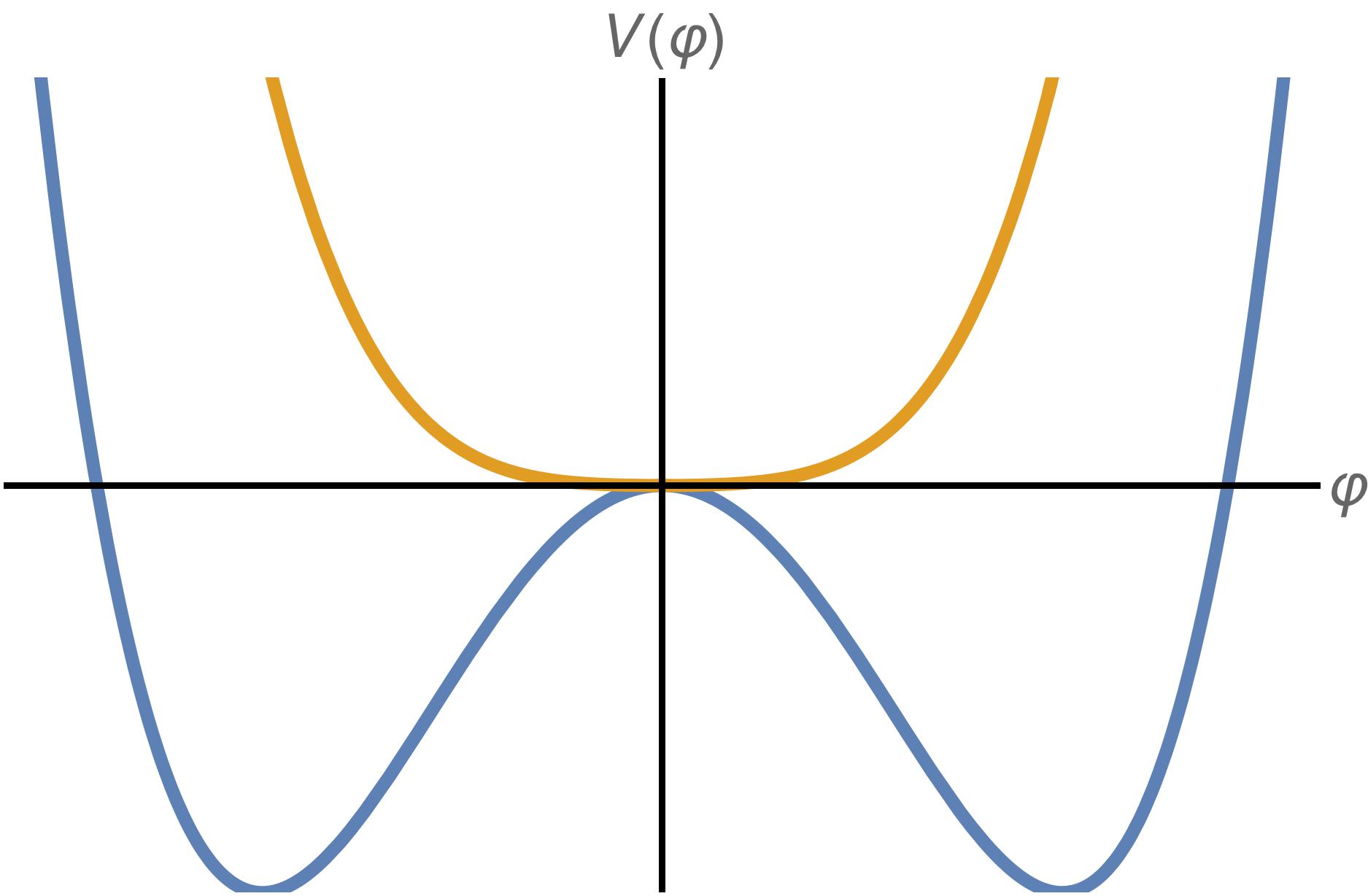}
\caption{At low values of the density $\rho$, the potential allows for spontaneous breaking of the $\mathbb{Z}_2$ symmetry and therefore gives a non-vanishing vaccum expectation 
value to the symmetron (blue line). However, for $\rho \gg \mu^2 M^2$, the symmetron can only have a vanishing vacuum expectation value and is therefore screened (orange line). }
\label{Fig:SymmetronPotential}
\end{center}
\end{figure}

The Lagrangian of the symmetron $\varphi$ is given by 
\begin{eqnarray}
\mathcal{L}_\varphi=-\frac{1}{2}\left(\partial\varphi\right)^2 -\frac{1}{2}\left( \frac{\rho}{M^2} - \mu^2 \right)\varphi^2 - \frac{\lambda}{4}\varphi^4,
\end{eqnarray}
where $\lambda$ and $M$ determine the strength of self-interaction and symmetron-matter coupling respectively. $\mu$ is a mass that, together with $M$ and the mass density $\rho$, controls the decoupling from matter. The effective potential in this Lagrangian has a $\mathbb{Z}_2$ symmetry (see Figure \ref{Fig:SymmetronPotential}) which can be spontaneously broken in environments of low mass density, such that the symmetron obtains a non-vanishing vacuum expectation value $\varphi_0$. However, in regions of high density, i.e. where $\rho \gg \mu^2 M^2$, the symmetry is restored and $\varphi$ can only take on a vanishing vacuum expectation value.

It is common practice to split a scalar field into a background value $\varphi_0$ and a small fluctuation $\delta\varphi$, such that $\varphi = \varphi_0 + \delta\varphi$. $\delta\varphi$ is the actual carrier of the fifth force and its interaction to matter is approximately proportional to $\rho\varphi_0\delta\varphi$ (at first order in $\delta\varphi$) which produces a force $F_\varphi \sim \varphi_0 \nabla\delta\varphi$. Consequently, the interaction to matter is turned off and the force is suppressed when the $\mathbb{Z}_2$ symmetry is restored in environments of large mass density since $\varphi_0 \equiv 0$ there.

For symmetrons we have $\zeta^2$ appearing in Eq. (\ref{eqn:conformaltrafo}) given by $\zeta^2\left(\varphi\right) = \varphi^2/2M^2$ and a field profile around a static spherically symmetric source given by \cite{Burrage2016}
\begin{eqnarray}\label{eqn:SymmetronProfile}
\varphi(r) = \varphi_{0,\text{out}} - (\varphi_{0,\text{out}} - \varphi_{0,\text{in}})\frac{R}{r}e^{m_{\text{out}}(R-r)c/\hbar}\frac{R m_{\text{in}}-
\tanh(m_{\text{in}} Rc/\hbar)}{R m_{\text{in}}+R m_{\text{out}}\tanh(m_{\text{in}} Rc/\hbar)},
\end{eqnarray}
where ``in'' and ``out'' denote quantities depending on the density in- and outside the sphere, respectively.

To simulate a symmetron field Jordan frame metric, the external and interaction potentials as defined in Eqs. (\ref{eq:lagrangian}) and (\ref{eq:interaction}) must be
\begin{equation}
\begin{split}V\left(r\right)=V_{0}&+\left[\frac{\partial_{r}\sqrt{\rho_{0}}}{\sqrt{\rho_{0}}}\left(\frac{1}{r}+\frac{m_{out}c}{\hbar}\right)-\frac{m_{out}^{2}c^{2}}{2\hbar^{2}}\right]\frac{\varphi\left(\varphi_{0,out}-\varphi\right)}{M^{2}}\\
 & \hphantom{+}+\frac{1}{2}\left(\frac{1}{r}+\frac{m_{out}c}{\hbar}\right)^{2}\left(\frac{\varphi_{0,out}-\varphi}{M}\right)^{2},
\end{split}
\end{equation}
\begin{equation}
\lambda=\lambda_0\left(1-\frac{\varphi^2}{2M^2}\right).
\end{equation}

\subsection{Dilaton}

Dilatons are scalar fields commonly appearing in discussions of string theory compactifications. Similar to symmetrons, their force is screened by a decoupling to matter in regions of large environmental density. In the string context their screening behaviour was first elaborated on in \cite{Damour1994}. A more modern discussion with implications for cosmology can be found in \cite{Brax20112}.

Following \cite{Joyce2014}, we write the dilaton Lagrangian as
\begin{eqnarray}\label{eqn:DilatonLagrangian}
\mathcal{L}_\varphi = -(\partial\varphi)^2 - V_0 e^{-\varphi/M_P}
- \frac{(\varphi - \varphi_*)^2}{2M^2}\rho,
\end{eqnarray}
where $V_0$ is a constant of mass dimension $4$, $M_P$ is the reduced Planck mass, $M$ controls the dilaton-matter coupling and $\varphi_*$ is a critical constant field value for which the dilaton decouples. This decoupling at $\varphi \approx \varphi_*$ is the essence of the dilaton screening mechanism since the value of $\varphi_*$ is choosen such that it occurs in environments of high mass density.

Here $\zeta^2$ appearing in Eq. (\ref{eqn:conformaltrafo}) is given by $\zeta^2\left(\varphi\right) = (\varphi - \varphi_*)^2/2 M^2 $ and if we expand the exponential potential in Eq. (\ref{eqn:DilatonLagrangian}) up to second order in $\varphi/M_P$, then the dilaton field profile is of the same form as the symmetron profile in Eq. (\ref{eqn:SymmetronProfile}).

To simulate a dilaton field Jordan frame metric, the external and interaction potentials as defined in Eqs. (\ref{eq:lagrangian}) and (\ref{eq:interaction}) must be
\begin{equation}
\begin{split}
V=V_{0}+\frac{1}{M^{2}}\left[\vphantom{\left(\frac{1}{2}\right)^2}\right.&\frac{1}{2}\left(\frac{1}{r}+\frac{m_{out}c}{\hbar}\right)^{2}\left(\varphi_{0,out}-\varphi\right)\\
& \left.\vphantom{\left(\frac{1}{2}\right)^2}+\left[\frac{\partial_{r}\sqrt{\rho_{0}}}{\sqrt{\rho_{0}}}\left(\frac{1}{r}+\frac{m_{out}c}{\hbar}\right)-\frac{m_{out}^{2}c^{2}}{2\hbar^{2}}\right]\left(\varphi-\varphi_{*}\right)\right]\left(\varphi_{0,out}-\varphi\right),
\end{split}
\end{equation}
\begin{equation}
\lambda=\lambda_0\left(1-\frac{\left(\varphi-\varphi_*\right)^2}{2M^2}\right).
\end{equation}

\subsection{Galileon}

Galileons were first introduced in the context of Dvali-Gabadadze-Porrati (DGP) braneworld models \cite{Dvali2000}. As described in more detail in \cite{Joyce2014}, galileons are higher derivative field theories with second order equations of motion. In (nearly) flat space the cubic galileon Lagrangian is \cite{Nicolis2008} 
\begin{eqnarray}\label{eqn:GalileonLagrangian}
\mathcal{L}_\varphi = -\frac{1}{2}(\nabla\varphi)^2 - \frac{1}{2\Lambda^3}\Box\varphi(\nabla\varphi)^2 - \frac{\varphi}{M}\rho,
\end{eqnarray}
where $\Box := \partial^\mu\partial_\mu$, $\Lambda$ controls the impact of the non-linear kinetic terms and $M$ determines the galileon-matter coupling. In principle, there are also quartic and quintic galileon terms (see e.g. \cite{Nicolis2008}) which are not included in this example. The free Lagrangian (including potential quartic and quintic terms) is symmetric under the galilean shift
\begin{eqnarray}
\varphi(x) \mapsto \varphi(x) + c +b_\mu x^\mu.
\end{eqnarray}
While the matter coupling term breaks this symmetry, $M\gg\Lambda$ so the symmetry breaking is mild (and discussed further in \cite{Joyce2014}).

The galileon force is screened close to a massive object by the Vainshtein mechanism \cite{Vainshtein1972}. In short, within a certain radius - the Vainshtein radius $R_v$ - around a massive object, the non-linear terms of Eq. (\ref{eqn:GalileonLagrangian}) dominate over and consequently suppress the kinetic term. This shortens the interaction range of the galileon and therefore weakens the resulting force. More details on this can inter alia be found in \cite{Joyce2014}. 

Here we have $\zeta^2$ appearing in Eq. (\ref{eqn:conformaltrafo}) given by $\zeta^2\left(\varphi\right)=\varphi/M $ as in the chameleon case and 
the field profile around a static spherically symmetric source is given by \cite{Bloomfield2014}
\begin{eqnarray}
\varphi(r) &=& \frac{\Lambda^3}{8}\bigg(r^2 \bigg[\sqrt{1 + \frac{R_v^3}{r^3}} -1 \bigg] 
+ 3\sqrt{R_v^3 R} \bigg[\sqrt{\frac{r}{R}} ~{}_2F_1\bigg(\frac{1}{6},\frac{1}{2};\frac{7}{6};-\frac{r^3}{R_v^3} \bigg)
\nonumber
\\
&\phantom{=}&
\phantom{\frac{\Lambda^3}{8}\bigg(r^2 \bigg[\sqrt{1 + \frac{R_v^3}{r^3}} -1 \bigg] + 3\sqrt{R_v^3 R}}
-{}_2F_1\bigg(\frac{1}{6},\frac{1}{2};\frac{7}{6};-\frac{R^3}{R_v^3} \bigg)\bigg]\bigg),
\end{eqnarray}
where ${}_2F_1$ denotes a Gaussian hypergeometric function and the Vainshtein radius $R_v$ is given by
\begin{eqnarray}
R_v = \left( \frac{8\rho R^3}{3M\Lambda^3} \right)^{1/3}.
\end{eqnarray}
Analytic solutions for field profiles around planar and cylindrical sources can be found in \cite{Bloomfield2014}.

To simulate a galileon field Jordan frame metric, the external and interaction potentials as defined in Eqs. (\ref{eq:lagrangian}) and (\ref{eq:interaction}) must be
\begin{equation}
\begin{split}
V=V_{0}+\frac{\partial_{r}\sqrt{\rho_{0}}}{\sqrt{\rho_{0}}}\frac{\partial_{r}\varphi}{M}+&\frac{\Lambda^{3}}{8M}\left\{ \vphantom{\frac{1}{\sqrt{\frac{r^3}{R_\nu}}}}\right.3\left(\sqrt{1+\frac{R_{\nu}^{3}}{r^{3}}}-1\right)\\
& +\frac{1}{4}\sqrt{\frac{R_{\nu}^{3}}{r}}\frac{1}{\sqrt{1+\frac{r^{3}}{R_{\nu}^{3}}}}\left[-3+\frac{3r}{4}-\frac{21}{2r}-\frac{R_{\nu}^{6}}{6r^{5}}+\frac{3\left(3-r\right)}{4\left(1+\frac{R_{\nu}^{3}}{r^{3}}\right)}\right]\\
& \hphantom{+}+\frac{1}{4}\sqrt{\frac{R_{\nu}^{3}}{r}}\left[3+\frac{9}{2r}+\frac{R_{\nu}^{6}}{6r^{5}}\right]{}_{2}F_{1}\left(\frac{1}{6},\frac{1}{2};\frac{7}{6};-\frac{r^{3}}{R_{\nu}^{3}}\right)\left.\vphantom{\frac{1}{\sqrt{\frac{r^3}{R_\nu}}}}\right\} ,
\end{split}
\end{equation}
\begin{equation}
\lambda=\lambda_0\left(1-\frac{\varphi}{M}\right).
\end{equation}

\subsection{D-BIon}

D-BIons are a type of screened scalar fields based on braneworld related Dirac-Born-Infeld (DBI) theories and, similarly to galileons, are screened by the Vainshtein mechanism \cite{Burrage2014}. However, instead of relying on non-linear terms in second derivatives, they are screened via non-linearities in first derivatives. Their matter-coupled Lagrangian is given by
\begin{eqnarray}
\mathcal{L}_\varphi = \Lambda^4 \sqrt{1- \frac{(\nabla\varphi)^2}{\Lambda^4}} - \frac{\varphi}{M}\rho
\end{eqnarray}
and leads to the following field profile around a static spherically symmetric source \cite{Bloomfield2014}: 
\begin{eqnarray}
\varphi(r) &=& \frac{\Lambda^2 R^3}{R_v^2}\bigg(\sqrt{1 + \frac{R_v^4}{R^4}} -1 \bigg) 
- \frac{\Lambda^2 R_v^2}{R} \bigg[\frac{R}{r} ~{}_2F_1\bigg(\frac{1}{4},\frac{1}{2};\frac{5}{4};-\frac{R_v^4}{r^4} \bigg)
\nonumber
\\
&\phantom{=}&
\phantom{\frac{\Lambda^2 R^3}{R_v^2}\bigg(\sqrt{1 + \frac{R_v^4}{r^4}} -1 \bigg) 
- \frac{\Lambda^2 R_v^2}{R}}
-{}_2F_1\bigg(\frac{1}{4},\frac{1}{2};\frac{5}{4};-\frac{R_v^4}{R^4} \bigg)\bigg]
\end{eqnarray}
with Vainshtein radius
\begin{eqnarray}
R_v = \sqrt{\frac{\rho R^3}{3M\Lambda^2}}.
\end{eqnarray}
As in the galileon case, $\Lambda$ controls the impact of the non-linear kinetic terms and $M$ determines the galileon-matter coupling. Analytic solutions for field profiles around planar and cylindrical sources can be found in \cite{Bloomfield2014}. Again, we have $\zeta^2$ appearing in Eq. (\ref{eqn:conformaltrafo}) given by $\zeta^2\left(\varphi\right)=\varphi/M $.

To simulate a D-BIon field Jordan frame metric, the external and interaction potentials as defined in Eqs. (\ref{eq:lagrangian}) and (\ref{eq:interaction}) must be
\begin{equation}
\begin{split}
V=V_{0}+\frac{\partial_{r}\sqrt{\rho_{0}}}{\sqrt{\rho_{0}}}\frac{\partial_{r}\varphi}{M}+&\frac{\Lambda^{2}}{8M}\left\{ \vphantom{\left(\frac{1}{\sqrt{\frac{r^3}{R_\nu}}}\right)}\right.\left(\frac{r^{2}}{R_{\nu}^{2}}\left[\frac{5}{R^{2}}-1\right]+\frac{r^{7}}{4R^{2}R_{\nu}^{6}}+\frac{2R_{\nu}^{2}}{R^{2}r^{2}\left(1+\frac{R_{\nu}^{4}}{r^{4}}\right)}\right)\frac{1}{\sqrt{1+\frac{R_{\nu}^{4}}{r^{4}}}}\\
& \hphantom{+}-\left(\frac{r^{2}}{R_{\nu}^{2}}\left[\frac{5}{R^{2}}-1\right]+\frac{r^{7}}{4R^{2}R_{\nu}^{6}}\right)F_{21}\left(\frac{1}{4},\frac{1}{2};\frac{5}{4};-\frac{R_{\nu}^{4}}{r^{4}}\right)\left.\vphantom{\left(\frac{1}{\sqrt{\frac{r^3}{R_\nu}}}\right)}\right\} ,
\end{split}
\end{equation}
\begin{equation}
\lambda=\lambda_0\left(1-\frac{\varphi}{M}\right).
\end{equation}

\section{Conclusion}
\label{sec:conclusion}

We have shown how to simulate a conformally coupled screened scalar field spacetime for quantum excitations of a BEC. By modulating the applied external potential $V$ and a magnetic field near a Feshbach resonance to modulate the inter-atomic interaction strength, the density can be varied in space without modifying the speed of sound, thus reproducing the desired metric. This simulation is applicable to any conformally coupled screened scalar field model when the scalar field is constant in time.

We find in this particular case that directly simulating a Jordan frame metric and simulating the effect of a conformally coupled screened scalar field are functionally equivalent. We also find that the reduction of this problem to $1+1$ dimensions results in any effect vanishing, so a full $3+1$ dimensional analysis is required. Finally, we give specific examples of chameleon, symmetron, dilaton, galileon and D-BIon fields and explicit implementations of these particular models for the field around a spherical mass in a vacuum chamber.

Constructing such a simulation will be useful in several applications. Firstly, computer simulation of the behaviour of matter under the influence of screened scalar fields is challenging due to the non-linearities of the models. Secondly, simulating conformally coupled screened scalar field models in a covariant formalism will allow us to test the metrological scheme of an upcoming proposal for a detector \cite{wip} constraining these models, as well as informing future work on the effect of screened scalar fields on particle creation, decoherence or other interesting quantum effects.

\begin{acknowledgments}
The authors thank Clare Burrage and Ben Thrussell for helpful comments. D.H. acknowledges funding from the Vienna Doctoral Program on Complex Quantum Systems (CoQuS) under the Austrian Science Fund (FWF) project code W 1210-N25. C.K. is supported by the School of Physics \& Astronomy of the University of Nottingham and via a Vice-Chancellor's Scholarship for Research Excellence. R.H. and I.F. would like to acknowledge that this project was made possible through the support of the Penrose Institute and the grant `Leaps in cosmology: gravitational wave detection with quantum systems' (No. 58745) from the John Templeton Foundation. The opinions expressed in this publication are those of the authors and do not necessarily reflect the views of the John Templeton Foundation.
\end{acknowledgments}

\end{document}